\documentclass{optica-article}

\journal{opticajournal} 

\articletype{Research Article}

\usepackage{lineno}

\begin{document}

\title{A Novel Differential Pathlength Factor Model for Near-Infrared Diffuse Optical Imaging}

\author{Kaiser Niknam,\authormark{1} Mannu Bardhan Paul,\authormark{2} and Mini Das\authormark{1,2,3,*}}

\address{\authormark{1}Department of Physics, University of Houston, Houston, TX, 77204, USA\\
\authormark{2}Department of Biomedical Engineering, University of Houston, Houston, TX, 77204, USA\\
\authormark{3}Department of Electrical and Computer Engineering, University of Houston, Houston, TX, 77204, USA}

\email{\authormark{*}mdas@uh.edu} 


\begin{abstract*}
	Continuous-wave near-infrared diffuse optical imaging (DOI) relies on accurate differential pathlength factors (DPFs) for quantitative chromophore estimation. Existing DPF definitions inherit formulation-dependent limitations that can introduce large errors in modified Beer--Lambert law (MBLL) analyses and associated inverse estimations. We propose two new DPF formulations: an MBLL-based formulation derived directly from photon-transport statistics and an experimentally practical inverse-distance model that closely reproduces the results of the more rigorous MBLL-based formulation. These models are derived based on extensive Monte Carlo simulations whereby we exclusively assign the source--detector-separation dependence of optical density (OD) to the Beer--Lambert term for a given object geometry, boundary configuration, and measurement setup. The additive term $G$ (widely referred to as the geometry-dependent term) in the MBLL is identified as the OD intercept in the limit of zero source--detector separation, thereby excluded of any source-detector separation dependence for a given geometry and object being imaged,  enabling the new formulations derived here. Our absolute MBLL-based DPF model shows the relevance of logarithmic aggregation of exponentially attenuated photon contributions in the measured OD. This differs from the arithmetic mean-pathlength treatment used in conventional mean-pathlength DPF formulations which is widely used. Our practically usable model, which we call an inverse-distance DPF model, can be used more readily for inverse solutions in imaging and is shown to be consistent with the aforementioned more rigorous model. We benchmark the two proposed DPF models against existing formulations using both the absolute and differential (where the term $G$ cancels) forms of the MBLL. The proposed models achieve errors below 20\% across broad optical conditions, whereas conventional DPFs can exceed 100\% error. The theoretical predictions are further validated using controlled phantom experiments, demonstrating improved quantitative accuracy in CW-NIR imaging.
\end{abstract*}

\section{Introduction}
Near-infrared (NIR) deep tissue imaging approaches have been extensively investigated as a complementary modality to radiological imaging, which often does not provide functional or molecular information. The utility of NIR imaging lies in the ability of near-infrared light to penetrate several centimeters into biological tissue, enabling real-time estimation of chromophore concentration changes—particularly those of oxyhemoglobin (HbO$_2$) and deoxyhemoglobin (HHb)~\cite{Scholkmann2014c,Delpy1988,Patterson1989,Sevick1991,Fantini1995}. Time-domain (TD) and frequency-domain (FD) NIR imaging methods provide quantitative access to tissue optical properties by exploiting temporal or phase-resolved measurements of light propagation~\cite{Essenpreis1993,Kohl1998,Pirovano2021,Duncan1995,Duncan1996,Matcher1995}. 
In contrast, continuous-wave near-infrared (CW-NIR) diffuse optical imaging systems measure only steady-state intensity attenuation and therefore offer a simpler, more portable, and more cost-effective alternative.

In CW-NIR imaging, changes in detected light intensities measured at multiple time points and across a range of source--detector separations (in either reflectance or transmission mode) are used to map relative changes in tissue absorption associated with physiological variations. As a result, this modality is widely used to monitor tissue oxygenation and hemodynamics in both research and clinical settings~\cite{Scheeren2012,Scholkmann2014,Hou2019}. It has found broad applications, including functional brain mapping~\cite{Zhang2024,Wang2025}, muscle oxygenation monitoring during exercise~\cite{Tuesta2022,CorralPerez2024}, and assessment of tissue viability under pathological conditions~\cite{Weingarten2010,Andersen2024}. In all of these methods, owing to the limited penetration depth of near-infrared photons (typically within the 650--900~nm optical window), reflectance--mode measurements are most commonly employed in practical imaging configurations. Photons detected at larger source--detector separations, on average, sample deeper tissue volumes, thereby providing increased sensitivity to subsurface physiological changes.

Inverse problems in all of these NIR deep tissue imaging require an estimation or approximation of the true pathlength (for varying source--detector separations). The true pathlength is longer than the actual separation owing to multiple scattering within the tissue and thus depends on tissue optical properties, geometry, and the source--detector separation. For media with higher scattering, the photons undergo a significantly higher number of scattering events before arriving at the detector either in transmission or reflectance mode of imaging. This scaling, or increase in the average photon pathlength relative to the actual linear source--detector distance, is referred to as the differential pathlength factor (DPF). The estimation or modeling of the DPF is critical in inverse problems related to NIR imaging.

The DPF concept was originally introduced by Delpy \emph{et al.} as the ratio of the mean photon pathlength to the source--detector separation~\cite{Delpy1988}. Since then, research on DPF estimation has centered on two major challenges: (1) accurately determining the true optical pathlength that photons travel in tissue, and (2) developing models of DPF that capture pathlength effects with sufficient accuracy for quantitative applications. To address the first challenge, a variety of experimental and computational approaches have been explored. Time-domain NIR spectroscopy employs ultrashort light pulses and measures photon time-of-flight distributions directly, providing detailed information on pathlength distributions across tissues and wavelengths~\cite{Essenpreis1993,Kohl1998,Pirovano2021}. Frequency-domain methods use intensity-modulated light and detect phase shifts and amplitude demodulation, from which the mean photon transit time—and thus the average pathlength—can be inferred in vivo~\cite{Duncan1995,Duncan1996,Matcher1995}. These techniques enable absolute calibration of photon pathlength and precise quantification of DPF, but they require considerably more complex instrumentation and come at a substantially higher cost. Consequently, time- and frequency-domain systems remain far less common than CW systems. However, CW instruments fundamentally measure only steady-state attenuation and therefore cannot independently determine photon pathlength~\cite{Talukdar2013}. As a result, CW-NIR imaging often relies on DPF values obtained from time- or frequency-domain measurements or from approximate models. Recent studies have suggested that wavelength-dependent changes in CW attenuation may provide a possible approach for estimating DPF using only CW systems~\cite{Hebden2026}. Uncertainties in DPF estimation using each of these methods can introduce errors in the recovered absorption coefficients. In particular, since absorption- and separation-dependent DPF formulations vary with both the local absorption coefficient and source–detector separation, errors in the estimated DPF are inherently spatially dependent. As a result, they may introduce spatial distortions in the reconstructed image rather than merely producing a uniform scaling error.

In addition to experimental techniques, computational approaches such as Monte Carlo photon-transport simulations~\cite{Jacques2022} and diffusion-equation models~\cite{Fantini1999} have been employed to estimate pathlength factors under controlled conditions~\cite{Arridge1992,vanderZee1992,Boas2001,Strangman2003,Das2006,Dehghani2009,Piao2015,Bhatt2016b,Kamran2018,Zohdi2018}. While these methods can predict photon pathlength distributions for specified optical properties and geometries, they rely on accurate knowledge of both tissue anatomy and optical coefficients. Consequently, inaccuracies in anatomical representation or in the assumed optical parameters (e.g., the use of generic head models or literature-averaged values) can cause simulation-based estimates to deviate substantially from actual \textit{in vivo} or \textit{in vitro} measurements.

The second challenge lies in integrating pathlength information into a practical DPF model for CW measurements. Early studies defined the DPF as the ratio of the mean optical pathlength to the source--detector separation~\cite{Delpy1988}. This definition is mathematically equivalent to the slope of absorption-induced changes in optical density with respect to the absorption coefficient, normalized by the source--detector separation~\cite{Sassaroli2004}. Despite this, in practice, the DPF is often approximated as a constant for source--detector separations exceeding 2.5\,cm~\cite{Ultman1991,vanderZee1992,Duncan1995,Chatterjee2018}. This assumption is motivated by its simplicity and by the fixed separation typically used in CW-NIR measurements, leading the distance dependence of the DPF to be neglected~\cite{Kohl2000,Gervain2011,Talukdar2013,Scholkmann2014,Alexander2015,Scholkmann2018,Barstow2019,Koirala2024}. However, extensive evidence indicates that the DPF is not invariant, but depends on tissue optical properties and photon pathlength characteristics~\cite{Kohl2000,Scholkmann2014,Scholkmann2018}. Approximating the DPF as a constant can therefore introduce substantial systematic errors in recovered chromophore concentrations, compromising quantitative accuracy~\cite{Talukdar2013,Alexander2015,Barstow2019,Koirala2024}.

These considerations motivate the present study. We systematically examine how different definitions of the DPF influence the accuracy of absorption coefficient and chromophore concentration estimation in CW-NIR measurements. Using Monte Carlo photon--transport simulations spanning a broad range of optical properties and source--detector separations, we evaluate several widely used DPF formulations and compare them directly with two alternative definitions introduced in this work. By applying each formulation to estimate the absorption coefficient in every simulation—where the ground-truth values are known—we assess the strengths and limitations of each approach. In addition to the simulation-based analysis, we experimentally confirm the validity of the proposed definitions using controlled phantom measurements. Our results demonstrate that the new models presented here yield highly accurate estimates across a broad range of optical properties and separations, with consistent agreement between simulation and experiment. Collectively, these findings establish a practical framework that enhances the quantitative reliability of CW-NIR imaging, bridging experimental simplicity with quantitative rigor and advancing its potential for robust functional and clinical applications.

\section{Methods}

\subsection{CW-NIR Imaging Models}

Quantitative interpretation of CW-NIR measurements is governed by the modified Beer–Lambert law (MBLL), which extends the classical Beer–Lambert law to account for photon scattering in turbid media~\cite{Delpy1988}. For a given source--detector separation $d$, chromophore concentration $C$, and wavelength $\lambda$, the measured optical density $\text{OD}_d$ (absorbance) can be expressed as:
\begin{equation}
	\text{OD}_d = - \ln\frac{I_d}{I_0} = \mu_a(C,\lambda) \cdot d \cdot \text{DPF}(C,d,\lambda) + G
	\label{eq:MBLL}
\end{equation}
where $I_0$ is the incident light intensity, $I_d$ is the detected intensity at a distance $d$ from the source, $\mu_a(C,\lambda)$ is the absorption coefficient, $G$ is a geometry-dependent factor, and $\mathrm{DPF}(C,d,\lambda)$ is the differential pathlength factor. The DPF accounts for the elongation of photon trajectories due to scattering, effectively scaling the geometric separation $d$ to an effective mean optical pathlength between source and detector. 

The additive term $G$ accounts for attenuation effects not explicitly represented by the Beer--Lambert absorption term and is commonly interpreted in prior literature as a geometry-dependent correction factor. In the conventional MBLL formulation, $G$ is commonly associated with the optical density in the limit of vanishing absorption, $\mu_a \rightarrow 0$. Accordingly, $G$ is treated as independent of changes in $\mu_a$ for a fixed measurement geometry and therefore cancels when changes in optical density arising from changes in absorption are considered. In the broader MBLL literature, the term ``geometry'' has been used in a general sense and is considered to include boundary conditions, medium shape and dimensions, refractive-index mismatch, photon escape losses, scattering properties, measurement coupling, and, in some formulations, source--detector separation~\cite{Anderson1967,Oshina2021,Piao2015}. The relative contributions of these individual factors are not explicitly quantified within the conventional MBLL representation considered here. In the formulation developed in this work, we define the additive term $G$ such that it does not contain explicit source--detector-separation dependence. This allows the explicit source--detector-separation-dependent variation in optical density to be assigned to the first term of the MBLL, which contains the DPF, while maintaining excellent agreement with both Monte Carlo simulations and experimental measurements.

Reported DPF values range from about 1 up to 16 (most commonly between 3 and 6 \cite{Talukdar2013}), depending on tissue type, wavelength, and measurement geometry~\cite{Delpy1988,Ferreira2007,Felix2013,Chatterjee2018,Koirala2024}. In general, the DPF increases with higher scattering coefficients ($\mu_s$). Although the absorption coefficient does not directly determine the mean photon pathlength, greater absorption preferentially attenuates longer photon trajectories, thereby shortening the effective pathlength and reducing the measured DPF. The absorption coefficient is related to the chromophore concentration through the molar extinction coefficient $\alpha(\lambda)$:
\begin{equation}
	\mu_a(C,\lambda) = \alpha(\lambda) \cdot C
	\label{eq:mua}
\end{equation}
Together, Eqs.~(\ref{eq:MBLL}) and~(\ref{eq:mua}) emphasize the role of DPF as a critical scaling factor that links measured optical attenuation to the underlying chromophore concentrations.

\subsection{Existing DPF Models}

The DPF was originally introduced as the weighted average of photon pathlengths~\cite{Delpy1988,Essenpreis1993,Duncan1995,Sassaroli2004}, formally defined as
\begin{equation}
	\mathrm{DPF}(d) = \frac{\langle s \rangle_{d}}{d}
	\label{eq:DPF_mean_original}
\end{equation}
where $\langle s \rangle_{d}$ denotes the expected pathlength of photons detected at a source--detector separation $d$. As expected, $\langle s \rangle_{d}$ increases with both source--detector separation and the scattering coefficient. A variety of studies have applied this formulation to estimate the DPF. For example, Duncan \emph{et al.} utilized phase-resolved spectroscopy to directly measure photon transit times and convert them into absolute pathlength values to establish reference DPFs for the adult head and calf~\cite{Duncan1995}. Similarly, Essenpreis \emph{et al.} employed time-resolved spectroscopy to characterize the wavelength dependence of the DPF~\cite{Essenpreis1993}.

Equation~\ref{eq:DPF_mean_original} has a mathematically equivalent formulation in which the DPF is defined as a slope-based quantity, namely, the derivative of absorption-induced changes in optical density with respect to the absorption coefficient, normalized by the source--detector separation~\cite{Delpy1988,Sassaroli2004},
\begin{equation}
	\mathrm{DPF}_{\mathrm{slope}}(d)
	=
	\frac{1}{d}
	\frac{\partial \mathrm{OD}_d}{\partial \mu_a},
	\label{eq:DPF_slope_def}
\end{equation}
where $\mathrm{OD}_d$ denotes the optical density measured at separation $d$. This \textbf{slope-based definition of DPF} has been widely adopted because it provides a theoretically grounded means of estimating the DPF directly from intensity measurements, thereby avoiding the need for explicit knowledge of individual photon trajectories. In analytical studies, Ultman and Piantadosi showed that the derivative of optical attenuation with respect to absorption captures the effective photon pathlength in diffusive media~\cite{Ultman1991}. Experimental validation was provided by Kohl \emph{et al.}, who applied the slope-based approach to cortical measurements and demonstrated close agreement between intensity-derived DPF estimates and values obtained using phase-resolved spectroscopy~\cite{Kohl2000}. Furthermore, computational investigations have employed this formulation to evaluate optical sensitivity in complex biological environments; for example, Alexander \emph{et al.} used Monte Carlo simulations to assess pathlength variations in a mouse model of Alzheimer’s disease~\cite{Alexander2015}, while Bhatt \emph{et al.} extended the framework to thick, heterogeneous tissues to account for spatial variations in optical properties~\cite{Bhatt2016b}.

In addition to the slope-based DPF, some studies have computed the mean pathlength by statistically aggregating photon trajectories within complex tissue geometries using the arithmetic mean, rather than the absorption-weighted mean in Eq.~\ref{eq:DPF_mean_original}, and proposed a \textbf{mean-pathlength DPF} defined as
\begin{equation}
	\mathrm{DPF}_{\mathrm{mean}}(d)
	=
	\frac{\bar{s}(d)}{d},
	\label{eq:DPF_mean}
\end{equation}
where $\bar{s}(d)$ denotes the arithmetic mean pathlength of photons detected at separation $d$. Notably, $\langle s \rangle_{d}$ in Eq.~\ref{eq:DPF_mean_original} converges to $\bar{s}(d)$ in Eq.~\ref{eq:DPF_mean} as $\mu_a \to 0$. As examples of the mean-pathlength definition of the DPF, Althobaiti employed Monte Carlo simulations to determine pathlength scaling for the multilayered architecture of human skin~\cite{Althobaiti2023}, while Chatterjee \emph{et al.} applied similar transport modeling to investigate pathlength variations in reflectance photoplethysmography~\cite{Chatterjee2018}.

Incorporating the slope-based definition into the diffusion-equation framework yields an analytical expression for the DPF in a semi-infinite slab with homogeneous absorption and scattering coefficients. This formulation, referred to as the \textbf{semi-infinite DPF}, is given by~\cite{Fantini1999,Boas2001,Felix2013}:

\begin{equation}
	\mathrm{DPF}_{\mathrm{seminf}} = \frac{1}{2}\sqrt{\frac{3\mu_s(1-g)}{\mu_a}} \cdot \left[1 - \frac{1}{1 + d \sqrt{3 \mu_s (1-g) \mu_a}}\right]
	\label{eq:DPF_semi_inf}
\end{equation}
where $g$ denotes the scattering anisotropy factor. Within this semi-infinite geometry, the DPF increases with the source--detector separation $d$ before approaching an asymptotic value. For typical tissue optical properties, this saturation occurs once $d$ exceeds approximately 0.5--0.7\,cm~\cite{Fantini1999}. Consequently, for separations larger than 2.5\,cm, the DPF is frequently approximated as invariant~\cite{vanderZee1992}. Based on this asymptotic behavior, a \textbf{constant DPF} has been proposed and remains widely adopted in practice. It is defined as the limiting case of Eq.~\ref{eq:DPF_semi_inf} as $d \to \infty$~\cite{Fantini1999,Ferreira2007,Felix2013,Alexander2015,Scholkmann2018,Chiarelli2019,Barstow2019}:
\begin{equation}
	\mathrm{DPF}_{\mathrm{constant}} = \frac{1}{2}\sqrt{\frac{3\mu_s(1-g)}{\mu_a}}
	\label{eq:DPF_constant}
\end{equation}

The constant DPF formulation is extensively employed to quantify chromophore concentrations in deep-tissue measurements. By treating the pathlength factor as invariant within this asymptotic regime, measured attenuation changes can be directly converted into chromophore concentration estimates without requiring real-time pathlength monitoring. Fantini \emph{et al.}, for example, applied this approach to assess cerebral oxygenation in neonates~\cite{Fantini1999}, while Chiarelli \emph{et al.} employed it in high-density adult recordings to mitigate hemoglobin cross-talk~\cite{Chiarelli2019}. Although this simplification is standard in skeletal muscle oximetry~\cite{Barstow2019}, its accuracy depends critically on the stability of tissue optical properties; indeed, Ferreira \emph{et al.} cautioned that assuming a fixed scattering coefficient—implicit in the constant DPF definition—can introduce errors during dynamic physiological conditions such as muscle contraction~\cite{Ferreira2007}.

\subsection{Computational Framework}
\label{sec:MC}

To evaluate the limitations of existing DPF formulations and to guide the development of improved alternatives, we performed a series of Monte Carlo photon-transport simulations spanning a broad range of absorption and scattering coefficients representative of biological tissue. For each condition, the simulated optical density was combined with each DPF formulation to estimate the absorption coefficient, whose ground-truth value was known \textit{a priori}. This framework enabled a direct and quantitative assessment of estimation errors across all DPF models.

\begin{figure}[!b]
	\centering
	\includegraphics[width=\linewidth]{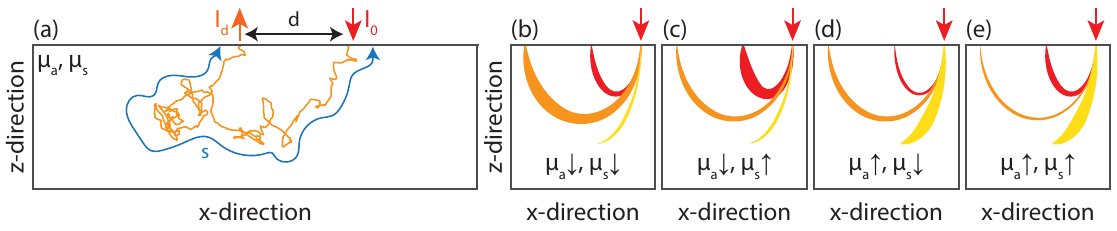}
	\caption{(a) Schematic of the Monte Carlo simulation. One million photons were launched into a homogeneous tissue slab using a pencil-beam source (red arrow), and their trajectories (orange) were recorded. The trajectory data were used to determine each photon's pathlength \(s\) and corresponding detected intensity, while the source--detector separation \(d\) was defined geometrically from the photon exit position (orange arrow) relative to the source (red arrow). 
		(b--e) Representative photon trajectories for different combinations of optical properties: 
		(b) low \(\mu_a\), low \(\mu_s\); 
		(c) low \(\mu_a\), high \(\mu_s\); 
		(d) high \(\mu_a\), low \(\mu_s\); and 
		(e) high \(\mu_a\), high \(\mu_s\). 
		Red, orange, and yellow traces denote photons that exited the top surface near the source (\(<1\,\mathrm{cm}\)), exited farther from the source (\(>1\,\mathrm{cm}\)), or were absorbed within the medium, respectively. 
		The thickness of each trajectory bundle is proportional to the number of photons following that path.}	\label{fig:Fig01}
\end{figure}

Simulations were performed using MCmatlab~\cite{Marti2018}, assuming a homogeneous tissue-slab geometry of \(29 \times 29 \times 5\)\,cm\(^3\). The slab dimensions were intentionally chosen to approximate a semi-infinite medium and to reduce boundary-reflection artifacts. In each simulation, approximately \(10^6\) photons were launched from a pencil-beam light source positioned at the center of the slab’s top surface (Fig.~\ref{fig:Fig01}a). This photon count was selected as a practical compromise between computational cost and accuracy owing to the large number of simulations required across the optical-property parameter space. All simulations were performed on the University of Houston Carya cluster using 1 CPU core and 64\,GB RAM, with typical runtimes of approximately 1--3\,h per simulation. Additional convergence tests comparing \(10^6\) and \(10^7\) photons for representative cases showed nearly identical results over the source--detector separation range of interest.

The absorption coefficient, $\mu_a$, was varied from 0 to 0.5\,cm$^{-1}$ in increments of 0.05\,cm$^{-1}$, while the scattering coefficient, $\mu_s$, ranged from 0 to 500\,cm$^{-1}$ in steps of 50\,cm$^{-1}$. The anisotropy factor $g$ was fixed at 0.95. As illustrated in the right panels of Fig.~\ref{fig:Fig01}, photon trajectories varied markedly with the medium’s optical properties. Increasing the scattering coefficient caused photons to exit closer to the source ($<\!1\,\mathrm{cm}$) owing to enhanced scattering, whereas lower $\mu_s$ values allowed photons to travel farther before exiting (compare Fig.~\ref{fig:Fig01}c with Fig.~\ref{fig:Fig01}b, and Fig.~\ref{fig:Fig01}e with Fig.~\ref{fig:Fig01}d). Similarly, higher absorption coefficients led to stronger attenuation, reducing the number of photons reaching the surface (compare Fig.~\ref{fig:Fig01}d\&e with Fig.~\ref{fig:Fig01}b\&c).

For each simulation, photon trajectories were recorded to determine the total pathlength traveled prior to absorption or escape from the medium. For photons exiting the top surface, additional quantities, including the exit location and exit weight, were stored. These data were used to compute the incident intensity ($I_0$), the detected intensity ($I_d$), and the corresponding optical density ($\mathrm{OD}_d$). The incident intensity was defined as the total launched optical power per unit area. 
The top surface of the medium was discretized into concentric radial bins of uniform thickness $\delta$. 
A central circular bin of radius $\delta$ was defined at the source location, and subsequent detection bins were defined as annuli spanning radii $[d, d+\delta]$, where $d = \delta, 2\delta, 3\delta, \ldots$
All exiting photons were assigned to bins based on their radial exit position relative to the source.

\subsubsection{New Empirical Model for Optical Density and $G$ Guided by MC Simulations}

Assuming the launch of $M$ photons (with $M = 10^6$ in this study), each initialized with unit weight and reference power per photon $P_0$, the incident intensity was computed as~\cite{Wang1995,Marti2018}
\begin{equation}
	I_0 = \frac{M P_0}{\pi \delta^2}
	\label{eq:I_0}
\end{equation}
where $\pi \delta^2$ corresponds to the area of the central circular bin associated with the normally incident pencil-beam source. The detected intensity at a source--detector separation $d$ was obtained by summing the exit powers of all photons escaping through the annular detection bin spanning radii $d$ to $d + \delta$, and normalizing by the corresponding annular area,
\begin{equation}
	I_d
	= \frac{\sum_{i=1}^{N_d} P_0 w_i}{\pi (\delta^2 + 2 \delta d)}
	\label{eq:I_d}
\end{equation}
where $w_i$ denotes the exit weight of photon $i$, and $N_d$ is the total number of photons exiting within the detection annulus. The optical density at distance $d$ was then calculated as
\begin{equation}
	\mathrm{OD}_d = -\ln\!\left( \frac{I_d}{I_0} \right)
	= -\ln\!\left( \frac{\delta}{M (\delta + 2 d)} \sum_{i=1}^{N_d} w_i \right)
	\label{eq:ODd}
\end{equation}

The discretization step \(\delta\) enters through the bin-area normalization applied to both the incident and detected intensities. For sufficiently small \(\delta\), variations in \(\delta\) primarily introduce a nearly uniform vertical offset in the optical density values while preserving the relative contrast and distance-dependent trends. Consequently, the reported OD values should not be interpreted directly as the logarithm of the fraction of detected photons relative to the total number of launched photons. Rather, they are computed from the sum of the detected photon weights after absorption and include the effects of the detector-area normalization. In this study, \(\delta \approx 2~\mathrm{mm}\). 

Furthermore, the weight of each photon attenuates exponentially with the total optical pathlength traveled within the absorbing medium. A photon $i$ that traverses a pathlength $s_i$ exits with weight

\begin{equation}
	w_i = \exp(-\mu_a s_i)
\end{equation}
where $\mu_a$ is the homogeneous absorption coefficient of the medium. The photon pathlength $s_i$ is determined by the scattering properties of the medium ($\mu_s$). Substituting this relation into Eq.~\ref{eq:ODd} yields
\begin{equation}
	\mathrm{OD}_d
	= -\ln\!\left(
	\frac{\delta}{M (\delta + 2 d)}
	\sum_{i=1}^{N_d} \exp(-\mu_a s_i)
	\right)
	\label{eq:ODd_mu}
\end{equation}

Finally, the geometry term $G$ is defined in this work as the optical density in the limit of zero source--detector separation:
\begin{equation}
	G = \mathrm{OD}_0
	\label{eq:G_term}
\end{equation}
This differs from the more conventional interpretation of $G$ as the residual optical density in the limit $\mu_a \rightarrow 0$~\cite{Oshina2021}. However, the same underlying logic applies in both cases in that $G$ represents the residual term when the Beer--Lambert attenuation contribution, $\mu_a \cdot d \cdot \mathrm{DPF}$, is driven to zero. In the existing literature or conventional interpretation, this occurs by taking $\mu_a \rightarrow 0$, whereas in the present formulation it occurs by taking $d \rightarrow 0$ such that the optical density is represented by the remaining additive term $G$. Consistent with the MBLL formulation in Eq.~\ref{eq:MBLL}, evaluating that equation at $d=0$ therefore directly yields Eq.~\ref{eq:G_term}.

The above-mentioned definition of $G$ as the optical-density intercept in the limit $d \rightarrow 0$ allows us to exclusively assign the source--detector separation dependence of the optical density to the first term $\mu_a \cdot d \cdot \mathrm{DPF}$. Thus, within the present framework, the geometry dependence of $G$ refers only to effects such as medium boundaries, shape, refractive-index structure, photon escape, and measurement configuration, rather than source--detector separation itself. In the subsequent sections, we will show that this new model characterizes optical density accurately while offering a pathway to simplify and improve accuracy for NIR reflectance inversion problems when measurements are made at multiple source--detector positions.

Substituting $d=0$ into Eq.~\ref{eq:ODd_mu} further gives
\begin{equation}
	G
	= -\ln\!\left(
	\frac{1}{M}
	\sum_{i=1}^{N_0} \exp(-\mu_a s_i)
	\right)
	\label{eq:G_MC}
\end{equation}
Such an analytical definition and experimentally approximable estimate of $G$ is necessary when the absolute form of the MBLL (Eq.~\ref{eq:MBLL}) is used, as is relevant to several imaging applications. In most CW-NIR imaging applications, the differential form of MBLL is used, where the term $G$ vanishes. In this work, we will examine the validity of the new DPF models presented here when both absolute and differential forms of MBLL are used. In the absolute form of MBLL, any new definition of DPF should be consistent with the expression for $G$. Since the conventional differential MBLL is derived directly from the absolute formulation, a DPF model that performs consistently within the absolute framework is also expected to remain applicable within the differential framework.

\begin{figure}[!t]
	\centering
	\includegraphics[width=\linewidth]{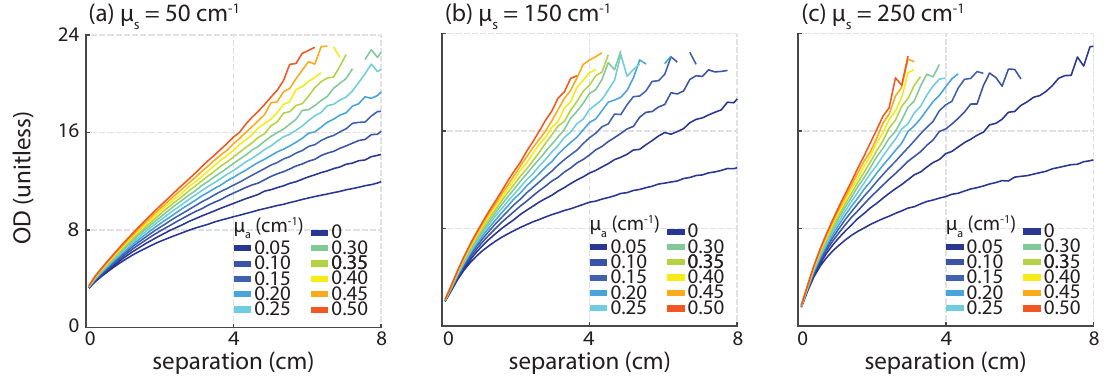}
	\caption{
		Optical density as a function of source--detector separation for absorption coefficients $\mu_a=0.05$--$0.50$ cm$^{-1}$ and three representative scattering coefficients: (a) $\mu_s=50$ cm$^{-1}$, (b) $\mu_s=150$ cm$^{-1}$, and (c) $\mu_s=250$ cm$^{-1}$. For each fixed scattering coefficient, the OD curves asymptotically converge to a narrow range of values as the separation approaches zero.
	}
	\label{fig:Fig01_5}
\end{figure}

We will show below that the expression for $G$ (Eq.~\ref{eq:G_term}) derived from Eq.~\ref{eq:MBLL} when $d=0$ is also consistent with results from our Monte Carlo simulations in which $\mu_a$, $\mu_s$, and source--detector separation were varied independently by changing one parameter at a time while holding the other two constant. Specifically, OD-versus-separation curves obtained for a fixed $\mu_s$ but varying $\mu_a$ exhibited intercepts that were substantially less sensitive to changes in $\mu_a$ than the corresponding OD values at larger separations. Representative examples are shown in Fig.~\ref{fig:Fig01_5}, where, for each fixed $\mu_s$, the OD curves converge toward a narrow range of values as $d \rightarrow 0$, despite significant variations in absorption coefficient. This weak absorption dependence at very small source--detector separations is consistent with the established physics of diffuse reflectance: when $\mu_s \gg \mu_a$, photons detected close to the source generally penetrate only shallow depths and travel relatively short pathlengths before returning to the surface, and therefore experience comparatively little absorption. In contrast, the intercept varies noticeably with $\mu_s$, suggesting a stronger dependence on scattering-related effects.

These observations provide additional support for defining $G$ as the optical density in the limit $d \rightarrow 0$. Under this definition, the intercept of the OD-versus-separation relationship is approximately independent of $\mu_a$ over the investigated optical-property range, despite the formal presence of $\mu_a$ in Eq.~\ref{eq:G_MC}, whereas the conventional interpretation of $G$ as the optical density in the limit $\mu_a \rightarrow 0$ generally retains an explicit dependence on source--detector separation. Thus, in the formulation adopted here, the explicit dependence on both $\mu_a$ and source--detector separation is absorbed into the Beer--Lambert attenuation term, while $G$ represents the remaining scattering- and geometry-related offset. These observations also motivate the development of the analytical DPF model presented in the following section.

\subsection{Proposed Novel DPF Models and Comparison with Existing Models}

Using the quantities obtained from the MC simulations, the various DPF formulations are computed for each combination of absorption and scattering coefficients $(\mu_a,\mu_s)$ in the simulated medium. We then compare the performance of the different DPF models using data from the same set of simulations.

\subsubsection{Computation of Existing DPF Models from Monte Carlo Simulations}

Before presenting our new DPF model, we briefly summarize how the quantities obtained from the MC simulations are used to compute the existing DPF models. Specifically, the \textbf{mean-pathlength} and \textbf{slope-based} DPFs defined in Eqs.~\ref{eq:DPF_mean} and~\ref{eq:DPF_slope_def}, respectively, are evaluated directly from the photon pathlengths $(s_i)$ recorded in the MC simulations. The \textbf{mean-pathlength} DPF is given by

\begin{equation}
	\mathrm{DPF}_{\mathrm{mean}}(d)
	=
	\frac{1}{d}
	\frac{\sum_{i=1}^{N_d} s_i}{N_d}
	\label{eq:DPF_mean_MC}
\end{equation}
Similarly, by combining Eqs.~\ref{eq:ODd_mu} and~\ref{eq:DPF_slope_def}, the \textbf{slope-based} DPF can be expressed as
\begin{equation}
	\mathrm{DPF}_{\mathrm{slope}}(d)
	=
	\frac{1}{d}
	\frac{\sum_{i=1}^{N_d} s_i \exp(-\mu_a s_i)}
	{\sum_{i=1}^{N_d} \exp(-\mu_a s_i)}
	\label{eq:DPF_slope}
\end{equation}
The \textbf{semi-infinite} and \textbf{constant} DPF models are computed using the absorption and scattering coefficients assigned to each simulation through Eqs.~\ref{eq:DPF_semi_inf} and~\ref{eq:DPF_constant}, respectively. This approach enables a direct comparison of all DPF formulations under identical optical conditions.

\subsubsection{Proposed MBLL-Based DPF Formulation}

While conventional DPF formulations, including the constant, semi-infinite, mean-pathlength, and slope-based models defined in Eqs.~\ref{eq:DPF_constant}, \ref{eq:DPF_semi_inf}, \ref{eq:DPF_mean}, and \ref{eq:DPF_slope_def}, are primarily derived from average photon pathlength estimates or diffusion-theory approximations, we propose an alternative formulation motivated by the absolute form of the MBLL and guided by our MC simulations. Rearranging Eq.~\ref{eq:MBLL} yields the following general expression for the DPF:

\begin{equation}
	\mathrm{DPF}_{\mathrm{MBLL}}(d)
	=
	\frac{1}{d} \cdot
	\frac{\mathrm{OD}_d - G}{\mu_a}
	\label{eq:DPF_true_original}
\end{equation}
Substituting our new formulation of optical density (Eq.~\ref{eq:ODd_mu}) (derived from MC simulations) into this relation yields the new model which we call henceforth \textbf{MBLL DPF}:
\begin{equation}
	\mathrm{DPF}_{\mathrm{MBLL}}(d)
	=
	-\frac{1}{d \cdot \mu_a}
	\left[
	\ln\!\left(
	\frac{\delta}{M(\delta + 2d)}
	\sum_{i=1}^{N_d} \exp(-\mu_a s_i)
	\right)
	+ G
	\right]
	\label{eq:DPF_true_formula}
\end{equation}
This formulation incorporates photon-transport statistics through the logarithmic aggregation of exponentially attenuated photon contributions, which differs from the arithmetic averaging employed in conventional mean-pathlength approximations. Although the MBLL DPF, defined directly from the full photon pathlength distribution, may be regarded as the most physically rigorous formulation within the adopted framework, its computation requires explicit knowledge of individual photon trajectories and is therefore computationally impractical for routine experimental or clinical applications. To bridge this gap between physical rigor and computational efficiency, we propose a new DPF model below.

\subsubsection{Proposed Analytical Inverse-Distance DPF Model}

Our new analytical formulation shown below is motivated by the observation that, although the dependence of optical density on source--detector separation $d$ is inherently nonlinear, it can be accurately approximated over physiologically relevant distances by a linear combination of $d$ and $\ln(d)$ with a nonzero intercept. This observation is also rooted in classical diffusion-theory analyses. In particular, Farrell \emph{et al.}~\cite{Farrell1992} and subsequent studies~\cite{Schmitt1990,Kienle1996,Piao2017,Chiarelli2019,Rudraiah2021} showed that, for a homogeneous semi-infinite scattering medium, the spatially resolved diffuse reflectance can be approximated as
\begin{equation}
	R = \frac{I_d}{I_0} \approx \frac{C_1}{d^{m}} \cdot \exp(-C_2 \cdot d)
\end{equation}
where the exponential term captures absorption-dominated attenuation and the power-law prefactor accounts for geometric photon spreading; $m$, $C_1$, and $C_2$ are medium-dependent constants determined by scattering and absorption. Taking the negative logarithm of the reflectance yields (see Eq.~\ref{eq:MBLL})
\begin{equation}
	\mathrm{OD}_d \approx C_2 \cdot d
	+ m \cdot \ln(d)
	- \ln(C_1)
	\label{eq:OD_Farrell}
\end{equation}
demonstrating that OD naturally decomposes into a linear term in $d$, a logarithmic correction, and a nonzero offset, as shown in Fig.~\ref{fig:Fig01_5}. Furthermore, as the absorption coefficient $\mu_a$ increases, the logarithmic contribution becomes progressively less significant relative to the linear term, and the OD curves increasingly approach a straight line with slope approximately equal to $C_2$ and a y-intercept approaching $-\ln(C_1)$, consistent with Eq.~\ref{eq:OD_Farrell}.

Monte Carlo simulations performed in this study (see Fig.~\ref{fig:Fig01_5}) confirm the validity of this approximation across a wide range of $\mu_a$ and $\mu_s$. For known or measured values of $\mathrm{OD}$, $G$, and $\mu_a$, the inverse-distance model parameters are obtained by fitting the quantity $(\mathrm{OD}_d-G)/\mu_a$ (which is equivalent to DPF) using ordinary least-squares linear regression in MATLAB (\texttt{fitlm}), with $d$ and $\ln(d)$ used as predictors together with an intercept term. Accordingly, the fitted model is
\begin{equation}
	\frac{\mathrm{OD}_d - G}{\mu_a}
	\approx
	A d + B + C \ln(d)
	\label{eq:OD_linear_fit}
\end{equation}
where, by comparison with Eq.~\ref{eq:OD_Farrell}, the fitting coefficients are given by $A=C_2/\mu_a$, $B=\left[-\ln(C_1)-G\right]/\mu_a$, and $C=m/\mu_a$. Because the model is linear in the fitted coefficients $A$, $B$, and $C$, no initial parameter estimates were required. The goodness of fit was quantified using the coefficient of determination ($R^2$).

\begin{figure}[!b]
	\centering
	\includegraphics[width=\linewidth]{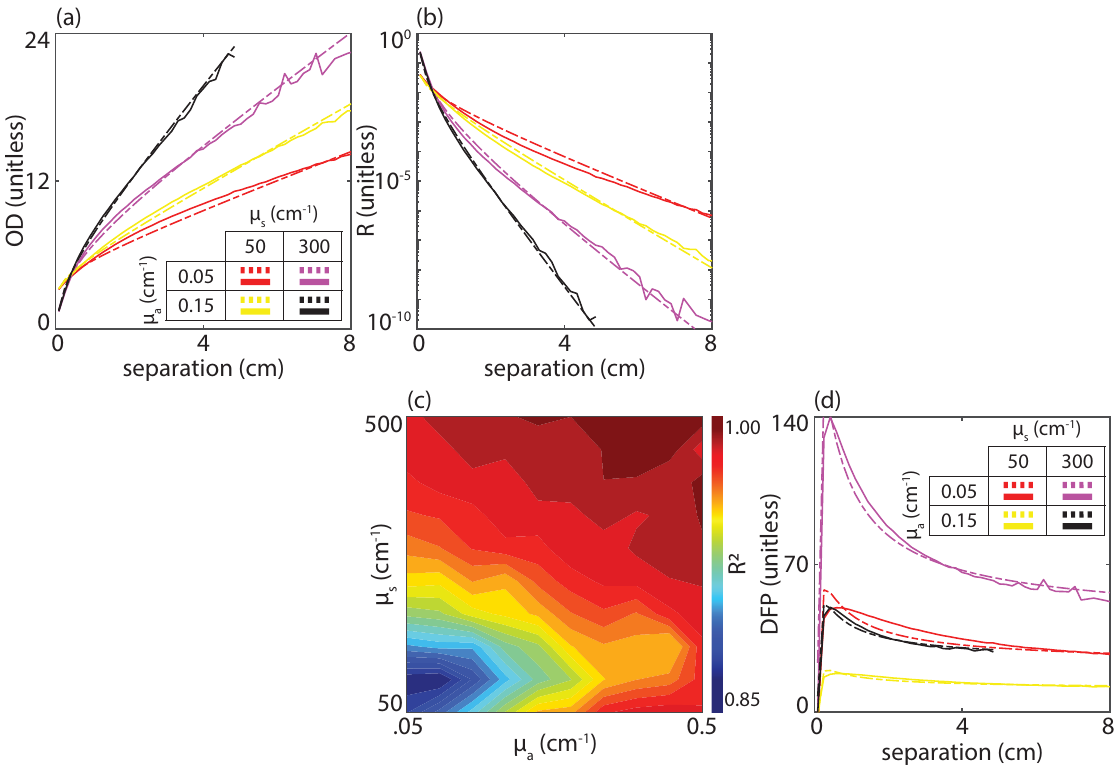}
	\caption{
		(a) Optical density as a function of source--detector separation for representative combinations of absorption and scattering coefficients $(\mu_a,\mu_s)$. Solid curves show Monte Carlo--derived OD values, while dashed curves represent fits of the same data to the mixed linear model given in Eq.~\ref{eq:OD_linear_fit}.
		(b) Corresponding diffuse reflectance as a function of source--detector separation for the same optical-property combinations shown in panel (a).
		(c) Heatmap of the ordinary $R^2$ values associated with the mixed linear fitting model across the simulated range of absorption and scattering coefficients.
		(d) Comparison of the corresponding MBLL DPF values (solid curves), computed directly from photon pathlength statistics, with the inverse-distance DPF estimates (dashed curves) obtained using Eq.~\ref{eq:DPF_inv}.
	}
	\label{fig:Fig02}
\end{figure}

Fig.~\ref{fig:Fig02}a shows representative OD profiles as a function of source--detector separation $d$ (solid curves) for multiple $(\mu_a,\mu_s)$ combinations. The corresponding normalized diffuse reflectance profiles ($R=I_d/I_0$) are shown in Fig.~\ref{fig:Fig02}b for the same optical-property combinations. When expressed in the mixed coordinate basis $\{d,\ln(d)\}$, the OD profiles are well approximated by linear regressions (dashed curves), from which the corresponding analytical approximations for the normalized diffuse reflectance profiles (dashed curves in Fig.~\ref{fig:Fig02}b) are obtained. The agreement between the simulated OD curves (solid lines) and their mixed linear approximations (dashed lines) is further reflected in the goodness-of-fit analysis shown in Fig.~\ref{fig:Fig02}c. Across the simulated range of $\mu_a$ and $\mu_s$, all retained fits yielded $R^2>0.85$, with an average value of $0.97 \pm 0.03$ (mean $\pm$ standard deviation). This high goodness of fit supports the use of the mixed linear representation as a compact surrogate model for the simulated OD profiles while preserving the essential transport behavior governing photon migration.

Based on these empirical and theoretical observations, we formalize the relationship into an explicit \textbf{novel inverse-distance DPF} model that retains analytical simplicity while improving absorption-recovery accuracy. Combining Eqs.~\ref{eq:DPF_true_original} and~\ref{eq:OD_linear_fit} yields
\begin{equation}
	\mathrm{DPF}_{\mathrm{inv}}(d)
	=
	A+\frac{B}{d}+\frac{C \cdot \ln(d)}{d}
	\label{eq:DPF_inv}
\end{equation}
Unlike the MBLL DPF, which must be recomputed for each discrete source--detector separation and optical-property combination using photon-transport simulations, the inverse-distance formulation provides a continuous analytical approximation across the spatial domain. As a result, once the parameters $A$, $B$, and $C$ are determined (see Results and Discussion section for the best approximations of these entities based on optical properties), the DPF can be estimated at arbitrary source--detector separations without requiring additional Monte Carlo simulations, thereby substantially reducing the computational cost.

We note, however, that practical estimation of the geometry-dependent term $G$ remains an important limitation of the absolute-form MBLL framework. In realistic settings, when optical density measurements are available at multiple source--detector separations and/or measurement conditions, $G$---assumed here to depend on the broader imaging geometry and boundary conditions, but not explicitly on source--detector separation---and the optical parameters may be estimated simultaneously by solving a coupled system of equations rather than treated as independently known quantities. Under such conditions, the need for additional simulations may be substantially reduced or eliminated.

Fig.~\ref{fig:Fig02}d compares the inverse-distance DPF values obtained from Eq.~\ref{eq:DPF_inv} with the corresponding MBLL DPF values computed directly from photon pathlength statistics. Across all optical-property combinations examined, the proposed model demonstrates close agreement with the MBLL DPF, capturing both its magnitude and distance dependence with minimal deviation. The agreement is particularly notable given that the inverse-distance formulation is derived from a compact analytical approximation requiring only three fitted parameters ($A$, $B$, and $C$), rather than explicit photon-pathlength information. It should be noted that the DPF values shown in Fig.~\ref{fig:Fig02}d are substantially larger than conventionally reported literature values because the proposed MBLL formulation is derived within a different theoretical framework than conventional pathlength-based DPF models. 

Furthermore, we demonstrate that $A$, $B$, and $C$ exhibit smooth and predictable dependencies on the absorption and scattering coefficients---and, by extension, chromophore concentrations---thereby enabling robust empirical calibration across a broad physiological range.

\subsubsection{Performance Evaluation of DPF Models}

The six DPF models (including our proposed MBLL and inverse-distance DPF models) considered here for comparison differ fundamentally in both their theoretical basis and computational complexity. The constant and semi-infinite formulations provide pragmatic but simplified approximations suitable for specific geometric regimes. The mean-pathlength and slope-based definitions are directly motivated by photon pathlength statistics, but they differ from the MBLL-based formulation in how these statistics enter the absorption-recovery model. The MBLL DPF offers the most rigorous and physically consistent formulation, albeit at substantial computational cost. Finally, the inverse-distance DPF introduced in this work achieves an optimal balance between accuracy and efficiency by approximating the MBLL DPF through a simple analytical expression.

Simulated optical density values were subsequently used to estimate the absorption coefficient using each of the six DPF models. Estimation accuracy was quantified across a broad range of source--detector separations and optical-property combinations by calculating the relative error
\begin{equation}
	e
	=
	\frac{\mu_{a,e}-\mu_a}
	{\mu_a}
	\times 100\%,
\end{equation}
where $\mu_a$ denotes the true absorption coefficient and $\mu_{a,e}$ is the estimated absorption coefficient obtained using a prescribed DPF as
\begin{equation}
	\mu_{a,e}
	=
	\frac{\mathrm{OD}_d - G}
	{\mathrm{DPF} \cdot d}
	\label{eq:mua_estimated}
\end{equation}
For absorption-coefficient-dependent DPF models, including the proposed MBLL formulation, Eq.~\ref{eq:mua_estimated} becomes implicit in $\mu_{a,e}$. In such cases, the estimated absorption coefficient was obtained as the value of $\mu$ that minimizes the relative residual of Eq.~\ref{eq:mua_estimated}, i.e.,
\begin{equation}
	\mu_{a,e}
	=
	\arg\min_{\mu}
	\left|
	1 - 
	\frac{
		\dfrac{\mathrm{OD}_d-G}
		{\mathrm{DPF}(\mu,d) \cdot d}
	}
	{\mu}
	\right|
	\label{eq:mua_implicit}
\end{equation}

Since conventional DPF formulations are traditionally defined with respect to a reference optical state, with absorption changes treated as perturbations about that state, the DPF term in Eq.~\ref{eq:mua_implicit} was evaluated at the true absorption coefficient, \(\mathrm{DPF}(\mu_a,d)\), rather than at the optimization variable, \(\mathrm{DPF}(\mu,d)\). This choice avoids introducing an iterative dependency that is not inherent to the original formulations and could otherwise bias their performance. Consequently, fixing the DPF at \(\mu_a\) provides a fairer assessment of the intrinsic accuracy of the DPF models and avoids confounding the comparison with errors introduced by the inversion procedure itself.

To assess the influence of the geometry-dependent term $G$ on the performance of the proposed DPF formulations, as well as their applicability within the differential form of the MBLL, an additional analysis was performed. In this analysis, the scattering coefficient of the simulated phantoms was held constant, while the absorption coefficient was varied. Assuming that $G$ is constant and independent of both $\mu_a$ and source--detector separation $d$, changes in OD resulting from changes in absorption were used to estimate the corresponding changes in $\mu_a$ using the various DPF models through Eq.~\ref{eq:MBLL}. Under this assumption, the differential form of the MBLL can be written as
\begin{equation}
	\Delta \mathrm{OD}_d
	=
	\mathrm{OD}_d
	-
	\mathrm{OD}_{d,0}
	=
	d
	\left[
	\mu_a \cdot \mathrm{DPF}(\mu_a,d)
	-
	\mu_{a,0} \cdot \mathrm{DPF}(\mu_{a,0},d)
	\right]
	\label{eq:diff_MBLL_general}
\end{equation}
which can be rearranged to yield the following expression for the change in the absorption coefficient:
\begin{equation}
	\Delta \mu_a
	=
	\mu_a-\mu_{a,0}
	=
	\frac{
		\displaystyle
		\frac{\Delta \mathrm{OD}_d}{d}
		+
		\mu_{a,0}
		\left[
		\mathrm{DPF}(\mu_{a,0},d)
		-
		\mathrm{DPF}(\mu_a,d)
		\right]
	}
	{\mathrm{DPF}(\mu_a,d)}
	\label{eq:diff_MBLL_general_3}
\end{equation}

In Eqs.~\ref{eq:diff_MBLL_general}--\ref{eq:diff_MBLL_general_3}, $\Delta\mathrm{OD}_d = \mathrm{OD}_d - \mathrm{OD}_{d,0}$ denotes the change in optical density at source--detector separation $d$, where $\mathrm{OD}_d$ and $\mathrm{OD}_{d,0}$ are the optical densities corresponding to absorption coefficients $\mu_a$ and $\mu_{a,0}$, respectively. The terms $\mathrm{DPF}(\mu_a,d)$ and $\mathrm{DPF}(\mu_{a,0},d)$ denote the differential pathlength factors associated with the two absorption states. For absorption-dependent DPF models, Eq.~\ref{eq:diff_MBLL_general_3} was solved numerically by minimizing the relative residual, as described in Eq.~\ref{eq:mua_implicit}. In contrast to the previous analysis, the DPF was not evaluated at the true absorption coefficient but was instead allowed to vary with the optimization variable during the minimization procedure. For DPF models assumed to be independent of absorption coefficient, Eq.~\ref{eq:diff_MBLL_general_3} reduces to the conventional differential MBLL formulation,
\begin{equation}
	\Delta \mu_a
	=
	\frac{\Delta \mathrm{OD}_d}
	{d \cdot \mathrm{DPF}(d)}
	\label{eq:diff_MBLL_general_4}
\end{equation}

The recovered changes in absorption coefficient were then scatter-plotted against the corresponding true values, and model performance was quantified using the mean magnitude of the relative errors, computed across all absorption coefficients at multiple source--detector separations. Because the geometry-dependent term $G$ cancels in the differential formulation, this analysis provides an independent assessment of DPF performance with substantially reduced dependence on the treatment of $G$. Consequently, it helps determine whether the observed performance differences are primarily attributable to the DPF formulation itself rather than to assumptions regarding the geometry term.

\subsection{Experimental Validation}

In addition to Monte Carlo simulations, experiments were conducted using a CW-NIR setup and tissue-mimicking phantoms to validate the findings in a physical setting. Two homogeneous scattering phantoms were prepared using a water-based super-absorbent polymer (SAP) hydrogel matrix (diaper polymer; Science Gone Fun, E-N Corp., Redford, MI, USA) mixed with titanium dioxide (TiO$_2$) powder. This composition was selected to provide a stable and repeatable scattering medium with relatively low intrinsic absorption, noting that the absorption of the hydrogel matrix is approximately that of pure water at near-infrared wavelengths, thereby enabling controlled manipulation of the optical properties. The phantoms were cast in an acrylic mold with dimensions $15 \times 15 \times 6$\,cm$^3$, providing a volume of approximately 1.35\,L to approximate a semi-infinite medium. The scattering agent consisted of TiO$_2$ powder (nominal particle size 44\,$\mu$m; Loudwolf Industrial \& Scientific, Dublin, CA, USA) dispersed in tap water. Both phantoms contained a SAP concentration of approximately 14\,g/L, with TiO$_2$ concentrations of 0.25\,mg/mL and 1.0\,mg/mL, respectively.

\begin{figure}[ht]
	\centering
	\includegraphics[width=\linewidth]{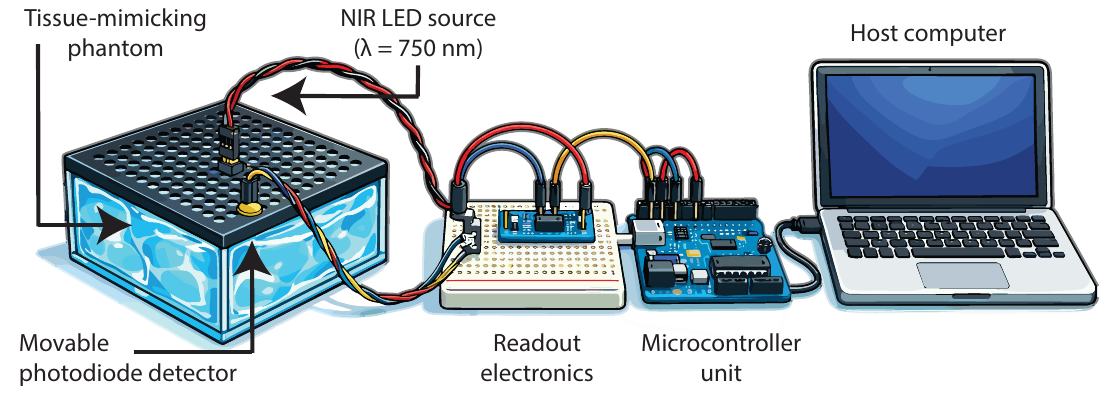}
	\caption{
		Experimental setup for validation using a homogeneous tissue-mimicking phantom. The phantom consists of a water-based super-absorbent polymer hydrogel matrix mixed with titanium dioxide particles and cast in a $15 \times 15 \times 6$\,cm$^3$ acrylic mold. A fixed 750\,nm light-emitting diode source illuminates the phantom surface at the center of a $13 \times 13$ measurement grid with a 1\,cm pitch, while a movable silicon photodiode detector samples diffuse reflectance at varying source--detector separations. The detected signal is conditioned by readout electronics, digitized by a microcontroller unit, and recorded on a host computer.
	}
	\label{fig:Fig03}
\end{figure}

The experimental setup utilized a light-emitting diode (LED) with a peak wavelength of 750\,nm (LED750L, Thorlabs, Newton, NJ, USA) to probe the phantom. The LED, housed in a TO-18 package with a spherical glass lens providing a narrow viewing angle ($\pm 11^{\circ}$), was driven by a stable 5\,V DC supply in series with a current-limiting resistor (150\,$\Omega$) to ensure constant optical output. Detection was performed using a silicon photodiode (FDS100, Thorlabs, Newton, NJ, USA) with an active area of 13\,mm$^2$, operating in a photovoltaic mode with a transimpedance amplifier circuit. The analog voltage signal, proportional to the detected light intensity, was digitized using an Arduino Mega 2560 microcontroller (Arduino, LLC) and recorded on a host computer (Fig.~\ref{fig:Fig03}).

Measurements were performed in a diffuse reflectance geometry. The source and detector were positioned on the phantom surface using a custom lid containing a \(13 \times 13\) grid of holes with a 1\,cm pitch (Fig.~\ref{fig:Fig03}). The lid was fabricated from matte black PLA using a fused deposition modeling (FDM) 3-D printer. The lower surface of the lid was therefore neither a perfect absorber nor a deliberately reflective optical coating, but rather a standard matte PLA surface. Additional controlled measurements performed with and without the lid showed differences in optical density comparable to the experimental noise level over the source--detector separation range considered in this work, indicating that the lid-induced boundary-condition effect was negligible within the experimental uncertainty. The LED was fixed at the center hole, and the detector was swept across the remaining grid positions. This configuration allowed the source--detector separation to be varied discretely from 1\,cm to approximately 8.5\,cm (center-to-center distance between the LED and the photodiode). For each separation distance, the optical density was calculated as $\mathrm{OD}_d = -\ln(I_d / I_0)$. The detected intensity $I_d$ was derived from the measured photodiode voltage, while the incident source intensity $I_0$ was estimated from the voltage--optical power characteristic curves of the light-emitting diode and independently confirmed through a controlled pre-measurement to ensure the validity of this theoretical assumption.

To evaluate the accuracy of the proposed differential pathlength factor models across different scattering regimes, reference optical properties---including absorption coefficient $\mu_a$, scattering coefficient $\mu_s$, and anisotropy factor $g$---were initially estimated from literature values reported for titanium-dioxide-based tissue-mimicking phantoms~\cite{Akarcay2012} corresponding to the nominal TiO$_2$ concentrations used (0.25\,mg/mL and 1.0\,mg/mL). Since the scattering properties of TiO$_2$ phantoms depend strongly on particle size, and the present study employed micron-scale rather than nanoparticle-scale TiO$_2$, the scattering properties were additionally refined using the experimentally measured optical-density-versus-separation curves. Table~\ref{tab:phantom_properties} summarizes the values obtained for these phantoms. These parameters were used to configure separate Monte Carlo simulations for each phantom, as detailed earlier. This approach served two purposes: first, to validate the experimental optical density trends against theoretical predictions for both concentrations, and second, to compute the theoretical mean optical pathlength of photons in each medium. The resulting pathlength data enabled the calculation of the standard DPF models (mean-pathlength, MBLL, etc.) as well as $G$. The computed DPF values, together with the corresponding estimates of $G$, were then substituted into Eq. \ref{eq:mua_estimated} using the experimental OD measurements to estimate the absorption coefficient and quantify the relative errors.
\begin{table}[!h]
	\centering
	\caption{Assigned optical properties of the homogeneous super-absorbent polymer--titanium dioxide phantoms at 750\,nm. Initial values were estimated from literature-reported optical properties of TiO$_2$-based tissue-mimicking phantoms~\cite{Akarcay2012} and further refined using the experimentally measured optical-density-versus-separation curves. The absorption coefficient $\mu_a$ primarily reflects the absorption of water at 750\,nm.}
	\label{tab:phantom_properties}
	\begin{tabular}{ccccc}
		\hline
		TiO$_2$ concentration & Refractive index & $\mu_a$ (cm$^{-1}$) & $\mu_s$ (cm$^{-1}$) & $g$ \\
		\hline
		0.25\,mg/mL & 1.33 & $0.0275$ & $13.75$ & $0.93$ \\
		1.00\,mg/mL & 1.33 & $0.0275$ & $55.0$ & $0.93$ \\
		\hline
	\end{tabular}
\end{table}
\section{Results \& Discussion}

\begin{figure}[!b]
	\centering
	\includegraphics[width=\linewidth]{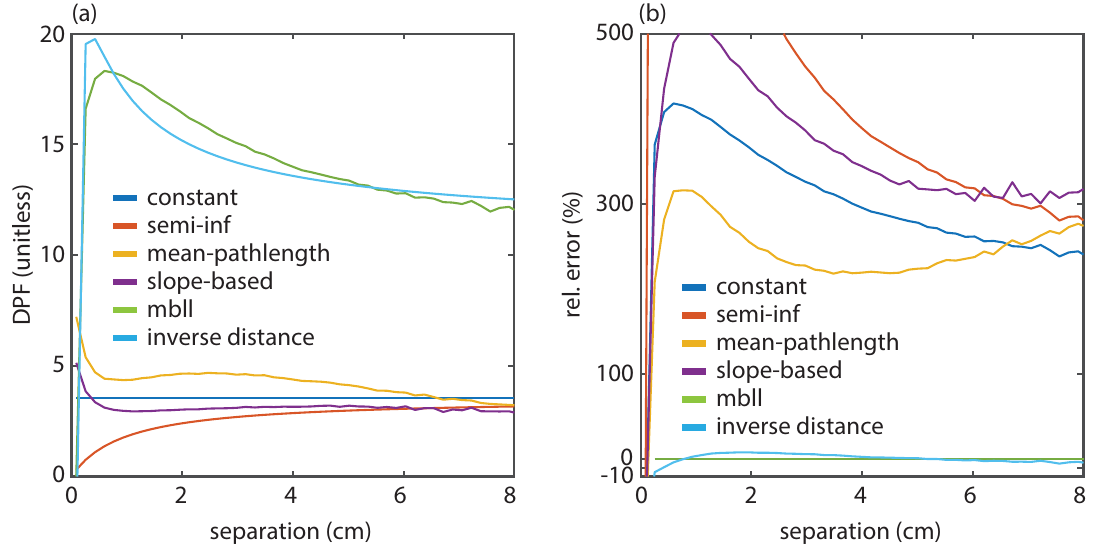}
	\caption{
		Comparison of DPF models and their corresponding absorption coefficient estimation errors for a representative optical property pair ($\mu_a = 0.15$\,cm$^{-1}$, $\mu_s = 50$\,cm$^{-1}$). 
		(a) Variation of the DPF as a function of source--detector separation for the constant, semi-infinite, mean-pathlength, slope-based, MBLL, and inverse-distance models. 
		(b) Relative error in estimating the absorption coefficient using each DPF formulation as a function of source--detector separation.
	}
	\label{fig:Fig04}
\end{figure}

Using the absolute form of the modified Beer--Lambert law, we systematically examined the behavior of the six DPF models across a wide range of source--detector separations and optical properties to evaluate their absorption-recovery performance. Fig.~\ref{fig:Fig04} (left panel) illustrates the variation of each DPF definition as a function of source--detector separation for a representative set of optical properties ($\mu_a = 0.15$\,cm$^{-1}$, $\mu_s = 50$\,cm$^{-1}$). The MBLL DPF and the inverse-distance DPF model exhibit remarkable agreement across most of the separation range, indicating that the proposed analytical approximation closely reproduces the pathlength statistics obtained from photon-resolved simulations. In contrast, the mean-pathlength, slope-based, and semi-infinite DPF formulations exhibit nearly identical behavior and, together with the constant DPF model, yield substantially lower DPF magnitudes than the MBLL formulation due to their different underlying definitions and formulations.

The right panel of Fig.~\ref{fig:Fig04} quantifies the corresponding relative error in estimating the absorption coefficient for each DPF model across source--detector separations ranging from 0 to 8\,cm. As expected, the MBLL DPF yields negligible error across all distances. The inverse-distance DPF model also demonstrates high accuracy, maintaining relative errors below 10\% over the majority of the separation range. Conversely, the constant, mean-pathlength, slope-based, and semi-infinite DPF formulations exhibit the poorest performance, producing substantial errors that frequently exceed 100\%. Notably, while the estimation errors associated with the mean-pathlength and slope-based DPFs do not exhibit a consistently decreasing trend with increasing separation, the constant and semi-infinite DPFs display a trend suggesting improved performance at very large separations ($>8$\,cm), a regime that lies beyond the practical operating range of many continuous-wave systems.

These trends persist across a broad parameter space, with absorption coefficients $\mu_a$ ranging from 0 to 0.5\,cm$^{-1}$ and scattering coefficients $\mu_s$ ranging from 0 to 500\,cm$^{-1}$. Across all optical property combinations examined, the MBLL DPF consistently yields zero estimation error, while the mean-pathlength, slope-based, semi-infinite, and constant DPF formulations produce errors exceeding 100\% at most source--detector separations.

\begin{figure}[!htbp]
	\centering
	\includegraphics[width=\linewidth]{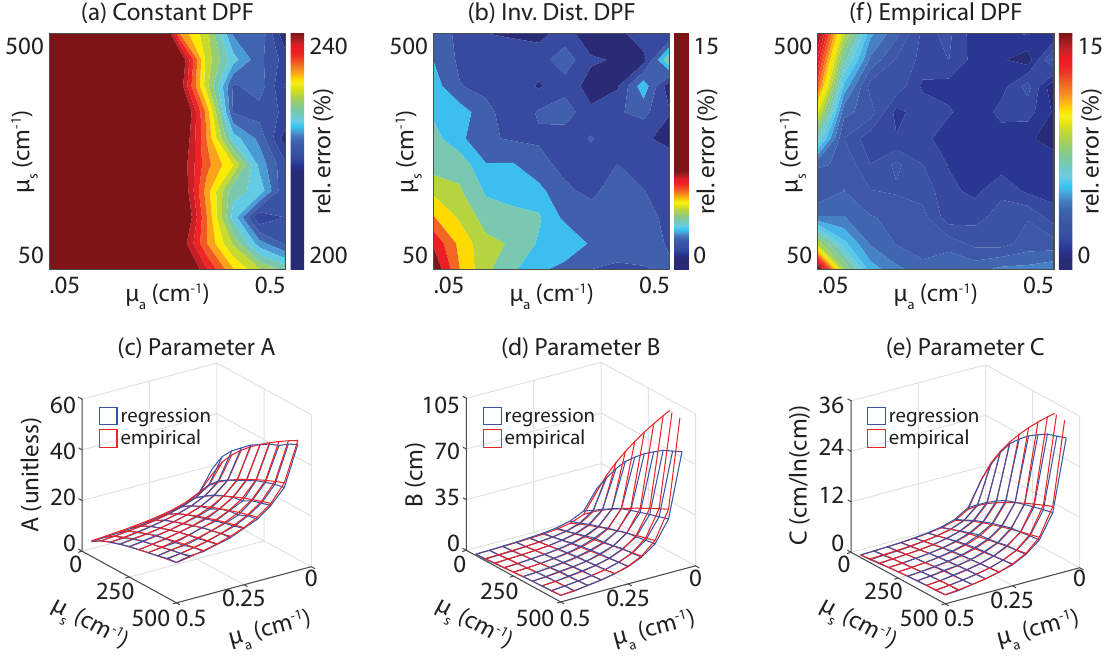}
	\caption{
		(a) Average magnitude of the relative error in estimating the absorption coefficient using the constant DPF model as a function of absorption and scattering coefficients. Errors are averaged over source--detector separations between 1 and 8\,cm. 
		(b) Average magnitude of the relative error obtained using the inverse-distance DPF under the same conditions. 
		(c--e) Variation of the inverse-distance DPF parameters $A$ (panel c), $B$ (panel d), and $C$ (panel e) across the $\mu_a$--$\mu_s$ space, comparing empirically fitted values (red) with regression-derived values (blue). 
		(f) Average magnitude of the relative error obtained using the fully empirical inverse-distance DPF model.}
	\label{fig:Fig05}
\end{figure}

Fig.~\ref{fig:Fig05}a depicts the magnitude of the relative error associated with the constant DPF model, shown here as a representative example of the previously proposed simplified DPF formulations. Errors are averaged over source--detector separations ranging from 1 to 8\,cm. This interval was selected to maximize clinical relevance: separations below 1\,cm provide insufficient depth sensitivity for deep-tissue interrogation, while separations exceeding 8\,cm generally surpass the effective detection limits of current CW-NIR systems. The results demonstrate that the constant DPF model maintains average errors exceeding 100\% across regions of lower absorption and scattering, which correspond to the typical optical properties of biological tissues~\cite{Jacques2013}. In contrast, Fig.~\ref{fig:Fig05}b depicts the magnitude of the relative error obtained using the inverse-distance DPF model, averaged over the same source--detector separation range. The inverse-distance model maintains an average error below 15\% across the majority of the optical property space. This level of accuracy is particularly robust in regions of lower absorption and scattering, which are most representative of biological tissues encountered in near-infrared spectroscopy applications.

Although the parameters $A$, $B$, and $C$ of the inverse-distance DPF model are obtained through linear regression, their dependence on the absorption and scattering coefficients is smooth and well behaved across the entire optical property space investigated (Fig.~\ref{fig:Fig05}c--e). Accordingly, the values of $A$, $B$, and $C$ obtained from each Monte Carlo simulation were further fitted over the simulated optical-property space using the power-law model $c\cdot \mu_a^{\,p} \cdot \mu_s^{\,q}$, where $c$, $p$, and $q$ denote fitting parameters. This procedure yields the following empirical relationships:

\begin{equation}
	A \approx 0.85 \times \mu_a^{-0.52} \times \mu_s^{0.41},
	\label{eq:eq_A}
\end{equation}
\begin{equation}
	B \approx 0.03 \times \mu_a^{-1.35} \times \mu_s^{0.65},
	\label{eq:eq_B}
\end{equation}
\begin{equation}
	C \approx 0.03 \times \mu_a^{-1.24} \times \mu_s^{0.54}.
	\label{eq:eq_C}
\end{equation}

These expressions offer additional insight and closely reproduce the regression-derived parameters, with coefficients of determination approaching unity for all three parameters. Fig.~\ref{fig:Fig05}c--e illustrates the comparison between empirically fitted and regression-derived values of $A$, $B$, and $C$ across the $\mu_a$--$\mu_s$ space, demonstrating excellent agreement. Using Eqs.~\ref{eq:DPF_inv}, \ref{eq:eq_A}, \ref{eq:eq_B}, and~\ref{eq:eq_C}, we constructed an empirical version of the inverse-distance DPF model. Fig.~\ref{fig:Fig05}f presents the corresponding average magnitude of the relative error, again averaged over source--detector separations between 1 and 8\,cm. As seen, the empirical model (Eqs.~\ref{eq:eq_A}--\ref{eq:eq_C}) closely reproduces the performance of the inverse-distance formulation (Eq.~\ref{eq:DPF_inv}) while eliminating the need for additional simulations or experimental calibration.

To evaluate whether the performance differences observed in the absolute-form MBLL analysis were influenced by the adopted treatment of the geometry-dependent term $G$, an additional analysis was performed using the differential form of the MBLL, in which $G$ cancels (Eq.~\ref{eq:diff_MBLL_general}). Monte Carlo simulations were performed as described above, with $\mu_a$ varied from 0 to 0.25~cm$^{-1}$ in increments of 0.01~cm$^{-1}$, while $\mu_s$ was held constant at 55~cm$^{-1}$. Equation~\ref{eq:diff_MBLL_general_3} was then used to estimate $\Delta\mu_a$. 

\begin{figure}[!b]
	\centering
	\includegraphics[width=\linewidth]{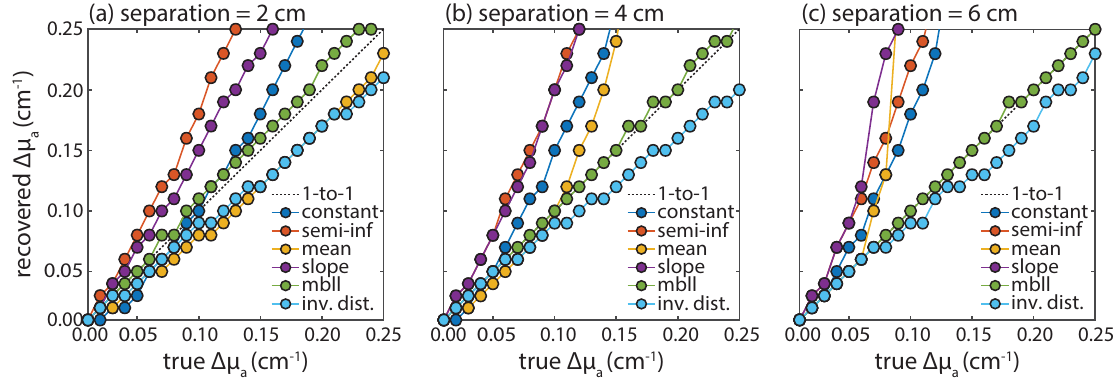}
	\caption{
		Recovered versus true changes in absorption coefficient obtained using the differential MBLL formulation for six DPF models at source--detector separations of (a) 2 cm, (b) 4 cm, and (c) 6 cm. The dotted line denotes the ideal one-to-one relationship.
	}
	\label{fig:Fig05_5}
\end{figure}

Fig.~\ref{fig:Fig05_5} summarizes the results of this analysis. For each source--detector separation, the recovered change in absorption coefficient was plotted against the corresponding true value. Across all source--detector separations examined, the MBLL DPF exhibited the closest agreement with the ideal one-to-one relationship, indicating the most accurate recovery of absorption changes. The inverse-distance formulation also demonstrated good agreement and consistently outperformed the constant, semi-infinite, mean-pathlength, and slope-based DPF models. Quantitative performance metrics are summarized in Table~\ref{tab:diff_mBLL_error}, confirming the visual trends observed in Fig.~\ref{fig:Fig05_5}. In particular, the MBLL DPF yielded the lowest error at all source--detector separations, while the inverse-distance formulation consistently produced the second-lowest error. Notably, the ranking of the DPF formulations remained unchanged from that observed in the absolute-form MBLL analysis. These results indicate that the improved performance of the MBLL and inverse-distance DPF models cannot be attributed solely to the treatment of $G$. Rather, it also arises from the ability of these formulations to more accurately recover absorption changes across the range of optical properties considered.

\begin{table*}[!t]
	\centering
	\caption{Mean magnitude of the relative error (\%) in recovered absorption coefficients obtained using the differential form of the MBLL for six DPF formulations at three representative source--detector separations. The lowest error in each column is shown in bold.}
	\label{tab:diff_mBLL_error}
	\begin{tabular*}{\textwidth}{@{\extracolsep{\fill}}lccc}
		\hline
		\textbf{DPF model} & \textbf{$d=2$ cm} & \textbf{$d=4$ cm} & \textbf{$d=6$ cm} \\
		\hline
		Constant & 27.62 & 61.85 & 97.68 \\
		Semi-infinite & 85.78 & 104.72 & 132.92 \\
		Mean-pathlength & 18.12 & 80.33 & 186.30 \\
		Slope-based & 48.96 & 130.23 & 206.53 \\
		MBLL & \textbf{5.87} & \textbf{1.18} & \textbf{0.22} \\
		Inverse-distance & 14.42 & 13.17 & 9.26 \\
		\hline
	\end{tabular*}
\end{table*}

\begin{figure}[!h]
	\centering
	\includegraphics[width=\linewidth]{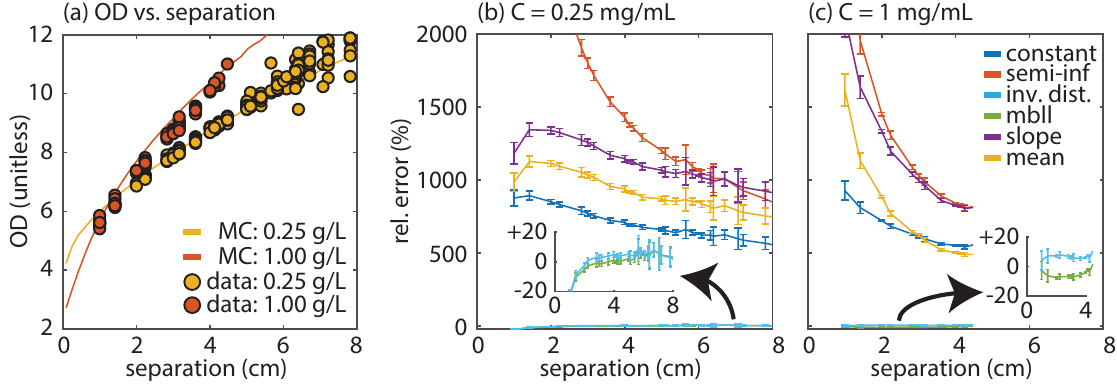}
	\caption{(a) Experimental validation of optical density as a function of source--detector separation. Measured optical density values (circles) for homogeneous phantoms with TiO$_2$ concentrations of 0.25\,mg/mL (yellow) and 1.00\,mg/mL (red) are compared with Monte Carlo predictions (solid lines). (b, c) Relative error in estimating the absorption coefficient for the (b) 0.25\,mg/mL and (c) 1.00\,mg/mL phantoms using six different DPF formulations. Error bars indicate the standard deviation of the estimation error. Insets show a magnified view of the low-error regime ($\pm 20\%$) to facilitate comparison among DPF models.}
	\label{fig:Fig06}
\end{figure}

To verify the simulation-based findings in a physical setting, we analyzed the diffuse reflectance data acquired from the two TiO$_2$ phantoms (0.25\,mg/mL and 1.00\,mg/mL). Fig.~\ref{fig:Fig06}a compares the experimentally measured optical density with Monte Carlo predictions initialized using literature-derived optical properties. The experimental measurements (circles) exhibit excellent agreement with the Monte Carlo results (solid lines) across the full range of source--detector separations for both phantom concentrations. This agreement is further quantified by coefficients of determination of 0.94 and 0.93 for concentrations of $C = 0.25$\,mg/mL and 1.00\,mg/mL, respectively. This close correspondence validates the experimental setup as well as the accuracy of the assumed optical properties, thereby establishing a reliable baseline for evaluating the performance of the DPF models. This validation step is essential, as accurate photon pathlength distributions—derived here from the validated simulations—are required to compute the MBLL, mean-pathlength, and slope-based DPFs.

Using the experimental optical density data, the absorption coefficients of both phantoms were subsequently estimated using the general MBLL inversion given by Eq.~\ref{eq:mua_estimated}, with the DPF term evaluated using each of the six formulations. Figs.~\ref{fig:Fig06}b and~\ref{fig:Fig06}c present the relative estimation errors for the 0.25\,mg/mL and 1.00\,mg/mL phantoms, respectively, as a function of source--detector separation. Consistent with the numerical results, the MBLL and inverse-distance DPF models demonstrate acceptable accuracy, maintaining relative errors below 20\% across the measured source--detector separation range. In contrast, the conventional DPF models, including the slope-based, semi-infinite, and constant formulations, exhibit pronounced inaccuracies, particularly at short separations, where estimation errors frequently exceed 100\%. Although these traditional models show moderate improvement at larger separations, they consistently underperform relative to the MBLL and inverse-distance DPF models.

Overall, these experimental findings corroborate the simulation-based conclusions and confirm that the MBLL and inverse-distance DPF formulations more accurately capture the separation-dependent photon pathlength scaling. Consequently, they provide a robust and physically consistent alternative to conventional DPF approximations when applied to realistic tissue-mimicking environments.

\section{Conclusion}

Accurate estimation of the differential pathlength factor is essential for reliable chromophore concentration measurements in CW-NIR diffuse optical imaging. Existing DPF definitions inherit formulation-dependent limitations that can introduce large errors in the MBLL analyses and associated inverse estimations. Here, we propose two new DPF formulations: an MBLL-based formulation derived directly from photon-transport statistics and an experimentally practical inverse-distance DPF model that closely tracks the more rigorous MBLL-based formulation. These models are derived based on extensive Monte Carlo simulations, from which we also identify the geometry-dependent term $G$ as the optical density intercept in the limit of zero source--detector separation. This formulation assigns the explicit source--detector-separation dependence of optical density to the Beer--Lambert term containing the DPF for a given object geometry, boundary configuration, and measurement setup, thereby enabling the new derivations presented here. Our absolute MBLL-based DPF model (Eq.~\ref{eq:DPF_true_formula}) shows the relevance of logarithmic aggregation of exponentially attenuated photon contributions in the measured OD. This differs from the arithmetic mean-pathlength treatment used in conventional mean-pathlength DPF formulations (Eq.~\ref{eq:DPF_mean_original}). Our practically usable model, which we call the inverse-distance model (Eq.~\ref{eq:DPF_inv}), can be used more readily for inverse solutions in imaging and is shown to be consistent with the more rigorous MBLL-based model.

In this study, we performed a systematic evaluation of four commonly used DPF formulations from the literature, namely the constant, semi-infinite, mean-pathlength, and slope-based models. We compared these formulations against our newly introduced MBLL DPF formulation and the practical inverse-distance DPF model proposed here with the goal of balancing absorption-recovery performance with computational efficiency.

Across a wide range of optical properties and source--detector separations, the inverse-distance DPF closely approximated the MBLL DPF and consistently provided more accurate absorption recovery than the conventional DPF formulations examined. Importantly, the qualitative ranking of the DPF models remained unchanged when evaluated using both the absolute and differential forms of the MBLL. The inverse-distance formulation offers several practical advantages. Once the model parameters are determined, the DPF can be evaluated continuously at arbitrary source--detector separations without requiring additional photon-transport simulations. Furthermore, because the model depends only on source--detector separation and a small set of fitted parameters, it can, in principle, be calibrated experimentally using CW-NIR measurements obtained at multiple separations. These details will become part of our future work. Although developed and validated within the continuous-wave framework, the proposed formulation is not restricted to CW-NIR systems and may be incorporated into time-domain or frequency-domain absorption-recovery models without modification of the underlying light-transport formalism.

The simulation-based findings were independently validated through controlled tissue-mimicking phantom experiments. The experimentally measured optical density trends and the corresponding absorption coefficient estimates closely matched the reference values, confirming that the observed performance differences among DPF models persist in a physical measurement setting. This agreement also supports the feasibility of experimentally estimating inverse-distance DPF parameters directly from CW measurements, further enhancing the practical relevance of the proposed formulation.

Our results highlight the critical importance of accurate DPF modeling, particularly at moderate source--detector separations where detected signal levels remain sufficiently high and measurement precision is greatest. In this regime, inaccuracies in conventional DPF models directly translate into substantial errors in absorption coefficient estimation, whereas the MBLL and inverse-distance formulations provide improved absorption-recovery accuracy. At larger source--detector separations, which are of particular interest in functional and clinical CW-NIR measurements for probing deeper tissue regions such as the cerebral cortex, some conventional DPF models exhibit modest improvements. However, the reduced signal-to-noise ratio at sufficiently large separations ultimately limits concentration-estimation accuracy regardless of the DPF formulation employed.

Overall, these findings underscore the practical utility of refined DPF formulations such as the inverse-distance model. While minor deviations from the MBLL DPF are observed at very short separations, these discrepancies are of limited practical concern due to the dominant sensitivity of such measurements to very superficial tissue layers. Consequently, the inverse-distance DPF provides a robust and efficient alternative to conventional models and is well suited for routine use in CW-NIR diffuse optical imaging, where rapid and reliable concentration estimation is required.

The inverse-distance formulation can be used to estimate absorption changes in experimental settings using OD measurements acquired at multiple source--detector separations. This approach requires an estimate of the geometry-dependent term $G$, which, according to our model, is equal to the OD at a source--detector separation of zero. In practice, $G$ may be approximated by measuring the OD as close to the source as experimentally feasible or by extrapolating OD measurements acquired at multiple source--detector separations. Our simulations indicate that, over a broad range of $\mu_a$ and $\mu_s$ values, approximating $G$ using the OD measured at source--detector separations as small as 0.5\,cm introduces only minor errors in the recovered absorption coefficient. Investigating the performance of the proposed DPF model under a broader range of experimental conditions and practical measurement strategies will be the subject of future work.

Future work will focus on extending this framework to heterogeneous tissue structures, for which the partial pathlength factor (PPF) provides a more appropriate metric than the DPF. In realistic clinical scenarios involving layered tissues or localized inclusions (e.g., tumors embedded within background tissue), the effective photon pathlength interacting with the inclusion depends on both optical contrast and source--detector separation. The distance-dependent formulation introduced here provides a natural foundation for modeling these effects and may be incorporated into inverse reconstruction schemes aimed at estimating inclusion-specific chromophore concentrations. It should also be noted that the DPF formulation presented in this work was derived for a semi-infinite homogeneous geometry. Consequently, its functional form is inherently geometry dependent and may require modification for other measurement geometries, boundary conditions, or tissue configurations where photon transport characteristics may differ from those considered here. An important open question concerns the extent to which DPF and PPF models can be generalized across different experimental configurations and application scenarios, including variations in tissue composition, geometry, and inclusion depth for concentration estimation. Addressing these questions may enable more accurate and robust quantification of tissue optical properties in both research and clinical applications, thereby enhancing the diagnostic and monitoring capabilities of diffuse optical imaging.

\begin{backmatter}
	
	\bmsection{Funding}
	National Institute of Biomedical Imaging and Bioengineering (R01 EB029761); Congressionally Directed Medical Research Programs (BC151607); National Science Foundation (1652892).
	
	\bmsection{Acknowledgments}
	The authors acknowledge the use of the Carya Cluster and the advanced support provided by the Research Computing Data Core at the University of Houston in carrying out the research presented in this study.
	
	\bmsection{Disclosures}
	The authors declare no conflicts of interest.
	
	\bmsection{Data availability}
	Data underlying the results presented in this study are not publicly available at this time but may be obtained from the authors upon reasonable request.
	
\end{backmatter}

\bibliography{sample}

\end{document}